\documentclass[10pt,aps,prb,twocolumn, nofootinbib]{revtex4-1}

\usepackage{amssymb,amsmath}
\usepackage{gensymb}
\usepackage{graphicx}
\usepackage{color}
\usepackage{hyperref}
\hypersetup{
	bookmarks=true,         
	unicode=false,          
	pdftoolbar=true,        
	pdfmenubar=true,        
	pdffitwindow=false,     
	pdfstartview={FitH},    
	pdfauthor={Kevin Multani},     
	colorlinks=true,       
	linkcolor=blue,          
	citecolor=red,        
}
\graphicspath{{figures/}}

\begin{document}
	
	\definecolor{dkgreen}{rgb}{0,0.6,0}
	\definecolor{gray}{rgb}{0.5,0.5,0.5}
	\definecolor{mauve}{rgb}{0.58,0,0.82}
	

	\title{Isotropic Huygens dipoles and multipoles with colloidal particles}

	\author{Romain Dezert}%
	\affiliation{University of Bordeaux, CNRS, CRPP-UPR8941,115 Avenue Schweitzer, F-33600 Pessac, France}%
	\author{Philippe Richetti}%
	\affiliation{University of Bordeaux, CNRS, CRPP-UPR8941,115 Avenue Schweitzer, F-33600 Pessac, France}%
	\author{Alexandre Baron}%
	\email{alexandre.baron@u-bordeaux.fr}
	\affiliation{University of Bordeaux, CNRS, CRPP-UPR8941,115 Avenue Schweitzer, F-33600 Pessac, France}%

	\date{\today}
	
	\begin{abstract}
Huygens sources are elements that scatter light in the forward direction as used in the Huygens-Fresnel principle. They have remained fictitious until recently when experimental systems have been fabricated. In this letter, we propose isotropic meta-atoms that act as Huygens sources. Using clusters of plasmonic or dielectric colloidal particles, Huygens dipoles that resonate at visible frequencies can be achieved with scattering cross-sections as high as 5 times the geometric cross-section of the particle surpassing anything achievable with a hypothetical simple spherical particle. Examples are given that predict extremely broadband scattering in the forward direction over a 1000 nm wavelength range at optical frequencies. These systems are important to the fields of nanoantennas, metamaterials and wave physics in general as well as any application that requires local control over the radiation properties of a system as in solar cells or bio-sensing. 
	\end{abstract}
	
	\pacs{Valid PACS appear here}
	\maketitle
	
	\noindent In 1690, Christiaan Huygens enunciated the famous principle that carries his name. It states that every point on a wave-front may be considered as a source of secondary spherical  wavelets which spread in the forward direction at the speed of light. The new wave front is the tangential surface to all of these secondary wavelets \cite{huygens1912}. Huygens sources can be obtained by overlapping the radiation from electric and magnetic dipoles that have equal polarizabilities\cite{geffrin_magnetic_2012}. The forward scattering arises from interferences between their emission patterns.
In the years 1980, this requirement of forward scattering was analyzed by Kerker et al. \cite{kerker1983electromagnetic} and the condition for forward scattering was given in the case of spherical particles. The ability to produce such sources with large dipole moments has important consequences because the stored potential energy inside the source can be large while maintaining a free-space matched forward radiation. As a result a net phase-delay can be imparted onto a wavefront with high efficiency, making it possible to make metasurfaces that act as optical surface components with 100\% transmission \cite{decker2015high,vo2014sub,arbabi2015subwavelength,yu2015high,decker2015high}, polarization beam controllers, splitters, converters and analysers \cite{xiang2016polarization,kruk2016invited} and perfect absorbers \cite{ra2015full}. The ability to produce such sources also has important implications for the fields of nanoantennas, where large scattering is desired with a well controlled radiation direction \cite{novotny2012principles,luk2015optimum}. Generally speaking, controlling the radiation of nanoparticles is essential to the design of both resonant and broadband anti-reflection coatings\cite{moreau2012controlled,akselrod2015large, spinelli2012broadband}.
     \begin{figure*}[t!]
		\centering
		\includegraphics[width=2 \columnwidth]{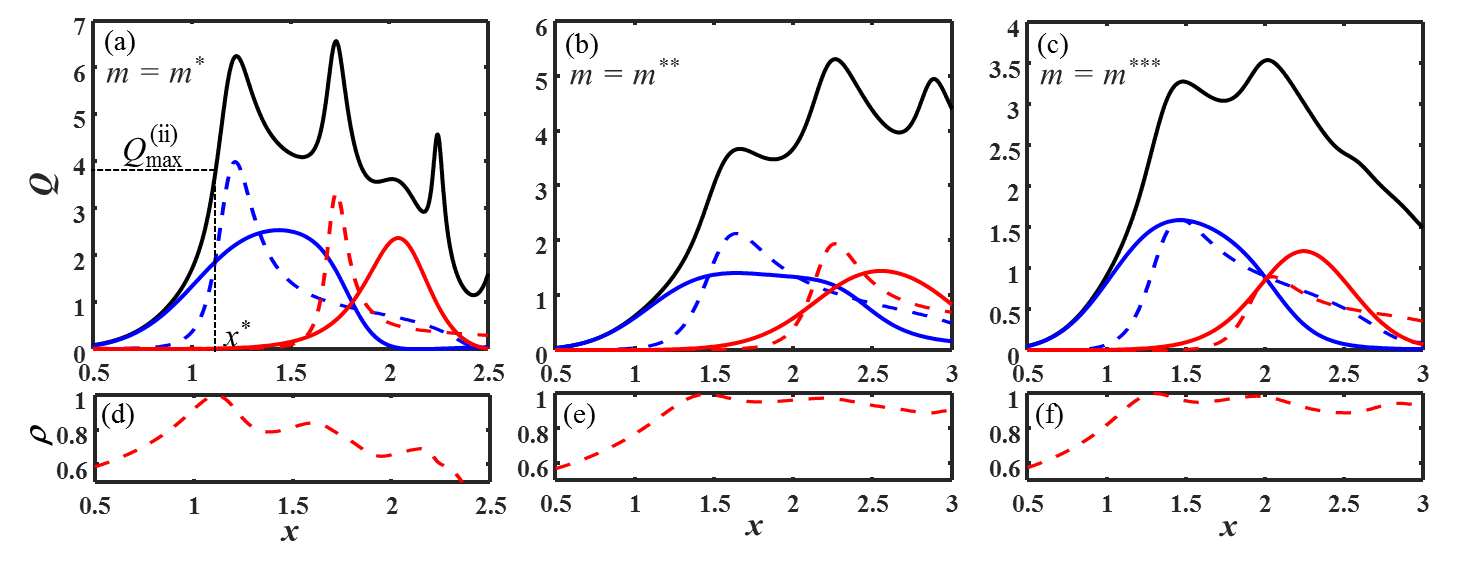}
		\caption{\textbf{\label{opt_diel} Three cases of scattering efficiency and forward to backward scattering ratio for a dielectric sphere.} The black solid curve represents the total scattering efficiency of the sphere as a function of the parameter $x$. The solid (dashed) blue curve is the scattering efficiency of the electric (magnetic) dipolar mode. The solid (dashed) red curve is the scattering efficiency of the electric (magnetic) quadrupole mode. (a) Scattering efficiency of a lossless dielectric sphere with $m = m^*$. The position of the optimal Kerker I condition (ii) is marked by the dashed black lines. (b) Scattering behavior of a dielectric sphere for which $m=m^{**}$, in which the electric dipole and magnetic dipole resonate at the same frequency but with different amplitudes. (c) Ideal Huygens dipole, when $m=m^{***}$ in which both the electric and magnetic dipole scattering resonate at the same frequency with equal amplitudes.(d-f) Fraction $\rho$ of energy scattered in the forward direction.}
	\end{figure*}
      
	We identify two classes of Huygens sources. The first kind are composed of anisotropic  particles - disks or cylinders - in which the magnetic $\sim \lambda/(2n)$ and electric $\sim \lambda/(n)$ modes resonate at the same wavelength $\cite{traviss2015antenna,lalanne2017metalenses}$. Both dielectric  \cite{decker2015high,kruk2016invited} and plasmonic systems have been proposed \cite{alaee2015generalized, rodriguez2014breaking}. 
    The second class of proposed Huygens sources are isotropic and composed of core shell nano-hybrids that overlap the electric dipole resonance of a plasmonic core with the magnetic Mie resonance of a high-index dielectric shell \cite{liu2014ultra}. These two kind of Huygens dipoles both present severe limitations. The anisotropic systems make the Huygens feature  highly dependent on the incident angle and an angle as small as 4$\degree$ reduces the transmission of a metasurface composed of such systems by almost 80\% for both \textit{s} an \textit{p}-polarizations \cite{lalanne2017metalenses}. In the end, this means that the metasurface optical component has a close-to-zero field of view. Alternatively the core-shell systems preserve their functionality for oblique incidences but at the expense of properties that critically depend on the geometrical parameters of the designs \cite{ra2015full}. 
	
	There is a dire need for experimentally accessible designs that act as efficient and isotropic Huygens sources either resonant or broadband. The recent progress of colloidal nanochemistry has made it possible to realize optical scatterers of both simple geometries containing a single material or more complex geometrical shapes composed of several materials that exhibit unusual properties \cite{ponsinet2015resonant,baron2016self,gomez2016hierarchical}. These bottom-up fabrication schemes are good candidates for the large-scale synthesis of Huygens sources at optical frequencies.
	
	The purpose of this article is to propose designs of efficient and brodband Huygens dipoles and multipoles based on colloidal systems. 
Let us first consider the best performances that may be reached by a lossless dielectric sphere of radius $a$ with a refractive index $n_p$. Let $m = n_p/n_h$ be the refractive index contrast, where $n_h$ is the host medium refractive index in which the sphere is immersed. We note $a_n(m,\mu,x)$ and $b_n(m,\mu,x)$ the $n^\mathrm{th}$ order Mie scattering coefficients of the electric and magnetic mutlipole that describe the behavior of the particle.
	They are both functions of the dimensionless parameter $x = 2\pi n_ha/\lambda$ and the magnetic permeability of the particle $\mu$. 
Whenever $a_n=b_n$ in the complex plane, the $n^{th}$ multipole will scatter forward. In the following, we will refer to this condition as \textit{Kerker I}. We may proceed to list three situations under which \textit{Kerker I} will be satisfied.
When $\mu=m$ (i), we have $a_n = b_n$. In the case where $\epsilon_\mathrm{h}=1$, \textit{Kerker I} simplifies to $\epsilon =\mu$, which means that the particle impedance is matched to that of the host medium, ensuring that any impinging wave perfectly couples to the particle \cite{kerker1983electromagnetic}. Since $\mu = 1$ at optical frequencies for all natural materials, this condition cannot be met with simple spherical particles.
When $mx = \psi'_{n,1}$ (ii), which is the first zero of the first derivative of the Ricatti-Bessel function $\psi_{n}$, we have $a_n=b_n$. 
When $mx = j_{n,1}$ (iii), which is the first zero of the spherical Bessel function $J_n$,
we have $a_n=b_n$. 
   Next we consider $Q^{(\mathrm{ii})}_n$ and $Q^{(\mathrm{iii})}_n$, the scattering efficiencies of the $n^{th}$ multipole under conditions (ii) and (iii) respectively, defined as the scattering cross-section of the $n^{th}$ multipole normalized to the geometrical cross-section of the sphere ($\pi a^2$). 
Condition (iii) occurs at higher frequencies than condition (ii) in a regime where several multipoles may interfere, so we shall focus on condition (ii) and note
 $Q^{(\mathrm{ii})}_{n,\mathrm{max}}$ the maximum achievable efficiency that satisfies Kerker I. Is is solely determined by $m$ (see supplemental materials). In the dipolar case, as already noticed by Luk'yanchuk et al. \cite{luk2015optimum}, the maximum efficiency reached is $Q_{1,\mathrm{max}}^{(ii)} \approx 3.72$, when $m = m^* \approx 2.455$, which means that $x = x* = \psi'_{n,1}/m^* \sim 1.118$. 
Figure \ref{opt_diel}(a) shows how the total efficiency evolves as a function of $x$ for a sphere when $m = m^*$. 
When $x =x^*$, the optimal efficiency is reached and the scattering efficiency of the electric and magnetic dipoles ($a_1$ and $b_1$) are equal, 
but this scattering does not occur at the resonance frequency of either dipole and the fraction of total scattered energy emitted in the forward direction $\rho$ peaks at $x^*$ and rapidly decreases as $x$ increases (see Fig. \ref{opt_diel}(d)).

We may then proceed to find a situation in which both the electric and magnetic dipoles resonate at the same frequency. This situation is given when $m=m^{**}\approx 1.87$. The scattering is shown on Fig. \ref{opt_diel}(b). We see that a maximum efficiency of 3.5 is reached for this dipole. Not all the scattering occurs in the forward direction because of a mismatch in amplitude between the electric and magnetic resonances, however a large fraction ($>$95\%) is still scattered forward above the first crossing. 

This situation can be considerably bettered by considering a lossy system. Indeed, since the magnetic dipole mode has a stronger interaction with the particle medium than does the electric dipole mode, losses decrease the scattering amplitude of the magnetic dipole and an optimum is reached when $m=m^{***} \approx 2.1+0.095i$ (see supplemental material), for which both amplitudes are equal and resonate at the same frequency (see Fig. \ref{opt_diel}(c)). In this case the maximum scattering efficiency reached is $\sim 3.2$, which is still almost 86\% of the value reached when $m=m^*$. We may refer to such dipoles as \textit{ideal Huygens dipoles}.
\begin{figure}[t!]
			\centering
			\includegraphics[width=1 \columnwidth]{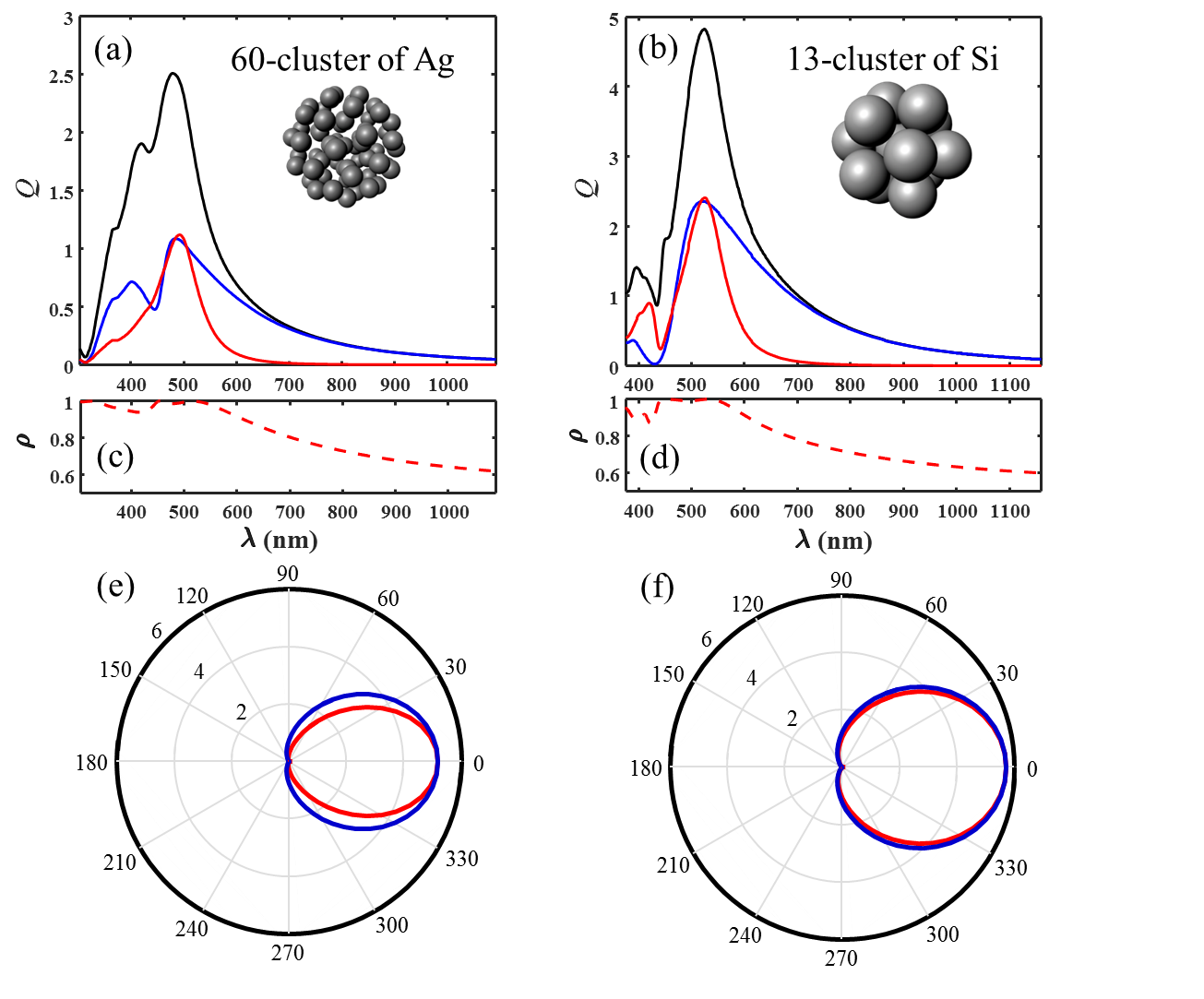}
			\caption{\textbf{\label{ultimate} Ideal Huygens sources with spherical clusters}. (a) and (b) are the scattering efficiencies as a function of wavelength for the 60-cluster made of Ag inclusions and the 13-cluster of silicon inclusions respectively. The black curve is the total efficiency. The blue (red) curve is the efficiency of the electric (magnetic) dipole. The insets are 3D renderings of what the clusters look like. (c) and (d) are the corresponding fractions of total energy scattered in the forward direction $\rho$. (e) and (f) are the far field radiation diagrams of the 60-cluster and the 13-cluster respectively. The blue (red) curve is the radiation diagram in the electric (magnetic) plane.}
		\end{figure}
Unfortunately it is hard to find materials that will reach $m^{***}$. 
A solution can be found by resorting to an effective medium sphere of radius $a$ made of spherical inclusions. Practically, this can be achieved by making a spherical cluster of nanospheres. In what follows, we shall consider clusters composed of identical inclusions of radius $a_i$ in repulsive interaction arranged by distributing them quasi-homogenously into the effective medium sphere of radius $a$. Fixing the values of $a_i$ and $a$ determines the range of the number $N$ of inclusions required to form the cluster. We will refer to it as an $N$-cluster and the volume fraction of inclusions in the cluster is $f=Na_i^3/a^3$. 
        
 \begin{figure*}[t!]
		\centering
		\includegraphics[width=2 \columnwidth]{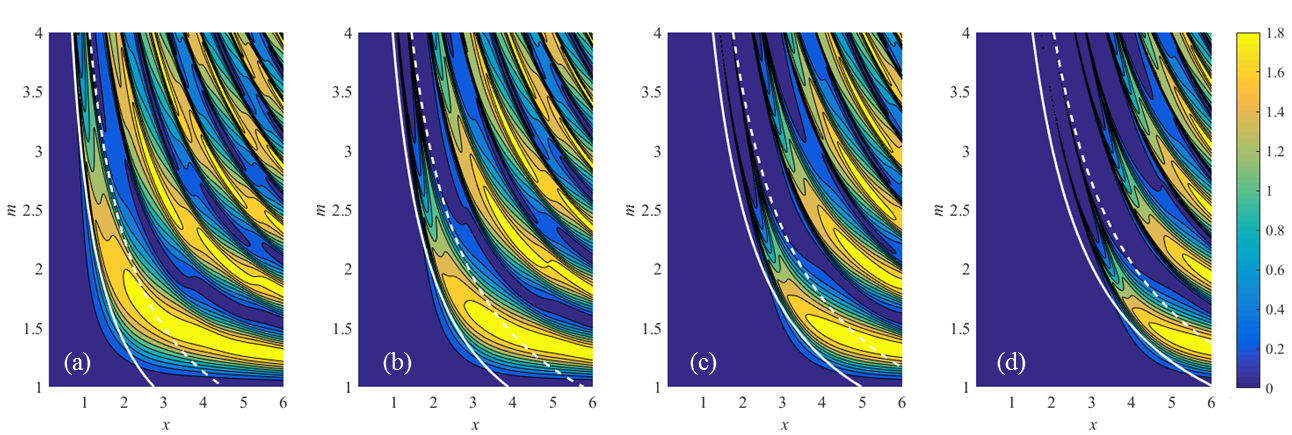}
		\caption{\textbf{\label{mp_traj} Multipolar trajectories.} Each panel is a contour map of $|a_n|^2+|b_n|^2$ for (a) $n=1$, (b) $n=2$, (c) $n=3$, and (d) $n=4$. The colormaps reveal the trajectories taken by the electric and magnetic multipoles in the $(m,x)$ plane. For any $n$, the white continuous curve traces the $mx=\psi'_{n,1}$ line, while the white dashed curve traces the $mx=j_{n,1}$ line. In other words the white lines are positions of which the Kerker I condition is fulfilled.}
\end{figure*}
    Targeting a peak resonance wavelength of $\lambda \approx 500$ nm using extended Maxwell-Garnett theory\cite{ruppin_evaluation_2000} (see supplemental material), we explored parameters $f$ and $a_i$ in the cases of silver and silicon inclusions with the aim of getting optical scatterers with overlapping electric and magnetic resonances at their peak. The optical properties of silver were taken from Palik's compendium \cite{palik1998handbook}, those of silicon were taken from Aspnes and Studna \cite{aspnes1983dielectric} and the scattering from the clusters were calculated using the T-matrix solver developped by Mackowski \cite{mackowski1996calculation,WinNT}, which uses Generalized Mie Theory and is rigorous in calculating the scattering of ensembles of spheres. Following this procedure we arrived at two spherical clusters that act as ideal Huygens sources. The first is a 60-cluster made of silver inclusions in water of radius $a_i = 15$ nm, with $f = 20\%$. The cluster radius is $a \approx 100$ nm and $m \approx 1.9 +0.18i$. The second is a 13-cluster made of silicon inclusions of radius $a_i = 41$ nm, with $f = 47\%$ in air. The cluster radius is $a \approx 123$ nm and $m \approx 2.0+0.13i$, as predicted by extended Maxwell-Garnett theory. 
    For both designs, the effective index contrast $m$ is close to $m^{***}$. Figures \ref{ultimate} (a,b) are the main result of this paper and show the total scattering efficiency of both systems. We see that the ideal Huygens dipole is reached in both cases near the designed wavelength. At the peak total scattering, both the electric and magnetic resonance overlap at the resonance wavelength of 500 nm (525 nm) for the silver cluster (silicon cluster) with equal amplitudes. The peak efficiencies reached are $\sim 2.5$ for the silver cluster and an impressive value of $\sim 5$ for the silicon cluster. Such a strong value for the the 13-cluster is a major result as the scattering efficiency reached by this Huygens dipole beats the theoretical maximum scattering efficiency reachable by an ordinary sphere with a complex refractive index following Mie theory as evidenced by Fig. \ref{opt_diel}(c). Furthermore this large response is unaccounted for by extended Maxwell-Garnett theory because $\mu_\mathrm{eff}$ is underestimated, while the real system is actually closer to condition (i). For each example, the amount of forward-scattered far-field energy relative to the total scattered energy is shown on Fig. \ref{ultimate}(c,d) and shows that the scattered energy is radiated in the forward direction over a broad range of frequencies around the dipole resonance. The radiation diagrams in both the electric (blue) and magnetic planes (red) almost perfectly overlap as shown on Fig. \ref{ultimate} (e,f). The results of the 13-cluster case are identical to those obtained using the commercial finite-element method solver COMSOL Multiphysics. 
\begin{figure}[b]
		\centering
		\includegraphics[width=1 \columnwidth]{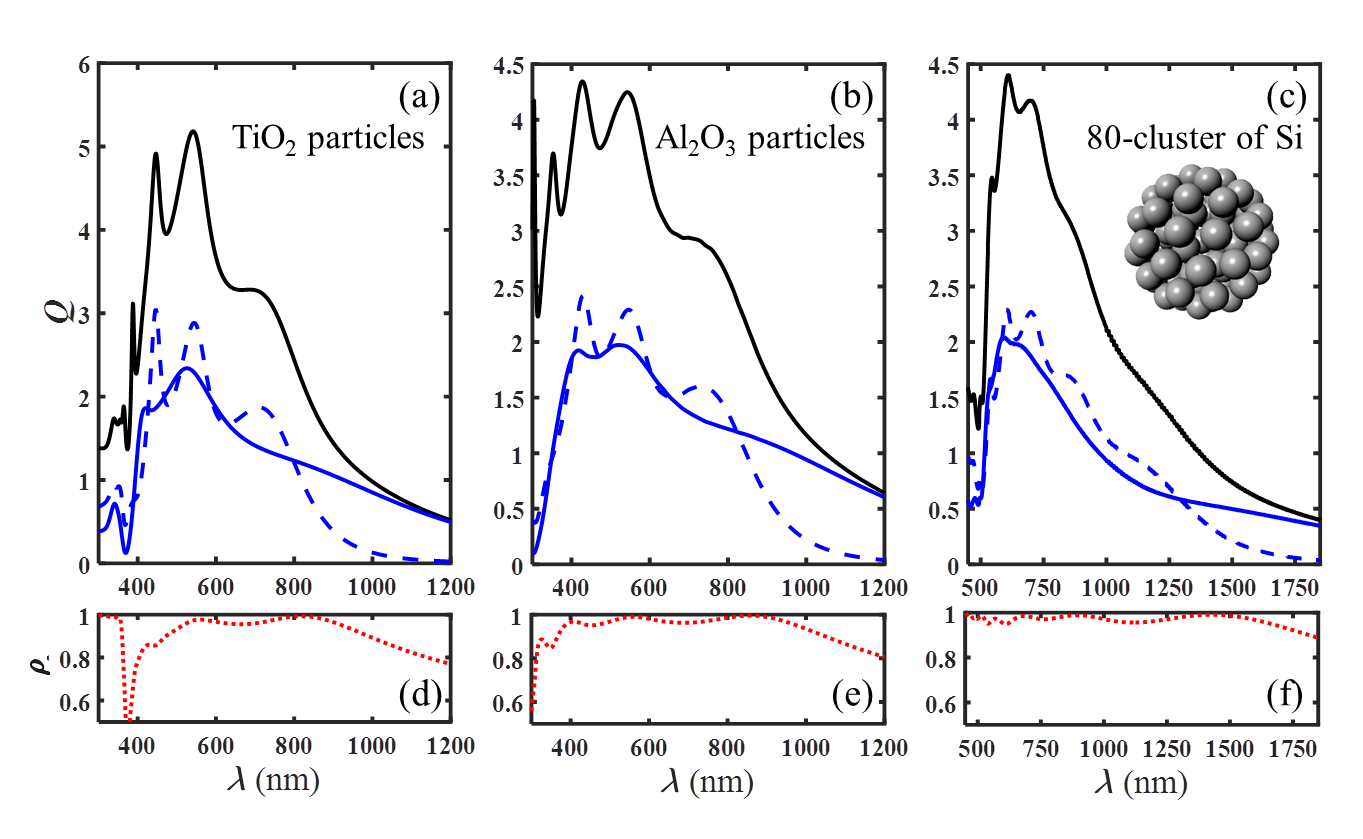}
		\caption{\textbf{\label{ex1} Extremely broadband scattering in the case for three examples of nanoparticles.} (a) Total scattering (black continuous curve) efficiency of a $TiO_2$ nanoparticle of 150 nm in radius in water ($n_h = 1.33$) for which $m \sim 1.5$. The total sum of scattering efficiencies by electric (magnetic) multipoles is represented as a blue continuous (dashed) curve. (b) ibid. for an $Al_2O_3$ nanoparticle of 200 nm in radius in air. (c) ibid for a 80-cluster of Si inclusions of 50 nm in radius in water and a volume fraction f = 37\% in water. The total radius of the cluster is 300 nm and $m$ given by Maxwell-Garnett theory varies from 1.4 to 1.6.}
	\end{figure}
	Up to now we have considered ideal Huygens dipole scattering. However, this condition  is resonant and only occurs over a narrow band around the resonance frequency. For numerous applications, it may be interesting to maximize the forward scattering over a broad range of wavelengths. Given a specific geometry and value of $m$, the Kerker I condition cannot be fulfilled exactly for all multipoles at once. But by relaxing the requirement that the electric multipole be exactly equal to the magnetic multipole and just satisfying the condition lossely, i.e. for all $n$, to obtain $a_n \sim b_n$, a broadband isotropic system may be designed. 
	
Indeed, by examining the trajectories taken by the scattering resonances of the electric ($a_n$) and magnetic ($b_n$) multipoles in the $(m,x)$ plane (see Fig. 3),we see that, while they remain well separated for large values of $m$, resonances all bundle together towards larger values of $x$ as $m$ becomes smaller than 2. This is apparent from the \textit{L} shapes taken by the $|a_n|^2+|b_n|^2$ shown on Fig. \ref{mp_traj}. 
The yellow area in Fig. \ref{mp_traj}, indicates multipoles satisfying the required condition that $|a_n|^2\sim|b_n|^2 \sim 1$, which means that multipoles have nearly equal amplitudes near their resonances. We see that the rigorous Kerker I condition under situations (ii), which are represented by the white dashed curves, even cross this merging area, when $m<2$.
Moreover for decreasing values of $m$, an increasing number of multipoles tend to overlap. In particular, for values close to 1.6,  the yellow area appears simultaneously for the first four multipoles. This means that the Kerker I condition can be approached multiple times and simultaneously for each couple of multipoles and forward scattering, though not perfect, will occur over a broad range of wavelengths.

	Such conditions can be met for colloidally synthesized nanoparticles such as $TiO_2$ or $Al_2O_3$, which are commercially available. Using empirical dispersions of the refractive index for both these compounds, we calculate the scattering efficiencies of a $TiO_2$ nanosphere of $a = 150$ nm immersed in water and of a $Al_2O_3$ nanosphere of $a = 200$ nm immersed in air in Fig. \ref{ex1}. The scattering by the electric and magnetic multipoles coincide over a broad range of wavelengths spanning over the interval [550 nm - 850 nm] with a scattering efficiency larger than 2 for $TiO_2$ and over the interval [400 nm - 1000 nm] with a scattering efficiency larger than 1 for $Al_2O_3$. The peak efficiency reached for $TiO_2$ ($Al_2O_3$) is $\sim 5$ ($\sim 4$). With a silicon cluster, the range of wavelengths can be considerably increased. Figure \ref{ex1}(c) shows that a broad Huygens multipole spanning over an impressive 1000 nm range can be achieved starting at 500 nm and for which $\rho$ is close to 1.  It should be noted that the condition we use that the electric and magnetic multipoles should overlap is loose and not rigorous. This condition is only used to get a broadband forward scattering system and not so much to explain the exact features of the forward scattering. Overall, this is satisfied as we see that the rough broad overlap between the scattering efficiencies of electric and magnetic multipoles coincides with a broad forward scattering ratio. However, it is not precise in determining the extrema of the forward scattering ratio nor the exact position of the cutoff. The spectral overlap of the efficiencies for a given wavelength is dominated by the overlap of a certain couple of electric and magnetic multipoles of order $n$, but is also slightly perturbed by the presence of non-overlapping multipoles of order $p \neq n$. For instance, this explains why the forward-scattering ratio in Fig. \ref{ex1}(e) reaches maxima at wavelengths slightly detuned compared to those where the electric and magnetic multipole efficiencies cross in Fig. \ref{ex1}(b).
   
   The interesting feature of these systems is that they are realistically achievable through colloidal self-assembly. Several recent examples show that it is possible to self-assemble plasmonic nanoparticles in clusters of submicrometric sizes\cite{dintinger2012bottom,schmitt2016formation,lacava2012nanoparticle}. The cluster system is rich as it is scalable and its optical properties can be tuned by varying the nature, amount, size and volume fraction of inclusions. Aside from these spherical clusters being experimentally feasible systems that produce isotropic Huygens sources, we believe they may serve as excellent building blocks for future nanoantennas and metasurfaces and inspire Huygens sources in other areas of wave physics such as acoustics or mechanics. Preliminary simulations suggest that the Huygens sources presented here can be used to produce phase-control metasurfaces. Several approaches exist to produce two-dimensional periodic surfaces with colloids as building blocks with impressive results \cite{yin2001template,xia2003template,cui2004integration,ni2015insights}. These approaches typically consist in coating suspensions of nanoparticles on patterned templates. Combining emulsion-based fabrication  of clusters and template-assisted self-assembly is a promising route to the fabrication of Huygens metasurfaces. 
   
\acknowledgments
	The authors acknowledge support from the LabEx AMADEus (ANR-10-LABX-42) in the framework of IdEx Bordeaux (ANR-10-IDEX-03-02), France.

	\bibliography{my_biblio2}

\end{document}